\documentclass[11pt,article,nofootinbib]{revtex4}
\usepackage{pdfpages}
\usepackage{epsf}
\usepackage{subfigure}
\usepackage{amsmath}
\usepackage{eqnarray,amsmath}
\usepackage{epstopdf}
\usepackage{mathrsfs}
\usepackage{natbib}
\usepackage{amssymb,latexsym}
\usepackage{dcolumn}
\usepackage{graphicx}
\usepackage{color} 

\makeatletter
\newcommand*{\rom}[1]{\expandafter\@slowromancap\romannumeral #1@}
\makeatother
\def\be{\begin{equation}}
    \def\ee{\end{equation}}
\def\ba{\begin{eqnarray}}
    \def\ea{\end{eqnarray}}

\graphicspath{{images/}}
\begin{document}
  \title{\large \bf Black hole shadow with a cosmological constant for cosmological observers }

\author{Javad \surname{T. Firouzjaee}$^{1}$}
\email{firouzjaee@kntu.ac.ir} 
\author{Alireza \surname{Allahyari}$^{2}$}
\email{alireza.al@ipm.ir}
\affiliation{ $^{1}$Department of Physics, K.N. Toosi University of Technology, Tehran 15875-4416, Iran, School of physics, Institute for Research in Fundamental Sciences (IPM), P. O. Box 19395-5531, Tehran, Iran\\
$^{2}$School of Astronomy, Institute for Research in Fundamental Sciences (IPM), P. O. Box 19395-5531, Tehran, Iran  }

\begin{abstract}

\textbf{Abstract:} We investigate the effect of the cosmological constant on the angular size of a black hole shadow. It is known that the accelerated expansion which is created by the cosmological constant changes the angular size of the black hole shadow for static observers. However, the shadow size must be calculated for the appropriate cosmological observes. We calculate the angular size of the shadow measured by cosmological comoving observers by projecting the shadow angle to this observer rest frame. We show that the shadow size tends to zero as the observer approaches the cosmological horizon. We estimate the angular size of the shadow for a typical supermassive black hole, e.g M87. It is found that the angular size of the shadow for cosmological observers and static observers is approximately the same at these scales of mass and distance.  We present a catalog of supermassive black holes and calculate the effect of the cosmological constant on their shadow size and find that the effect could be $3\; precent$ for distant known sources like the Phoenix Cluster supermassive black hole.

\end{abstract}
%
%\pacs{}
%
    %
    \maketitle
    
%%%%%%%%%%%%%%%%%%%
\section{Introduction}
%%%%%%%%%%%%%%%%%%%

Recent observations of black holes shadow at a wavelength of 1.3 mm spectrum succeeded in observing the first image of the black hole in the center of the galaxy M87 \cite{Akiyama:2019bqs,Akiyama:2019cqa}. Although the first image of the black hole does not distinguish the black hole geometry precisely, the future improvement of measurements will lead to a much higher resolution. This will allow us to study different aspects of cosmological black holes \cite{cbh}. To this end, one needs to calculate the influence of different parameters on the shape of cosmological (astrophysical) black holes shadow, e.g accretion plasma effects \cite{plasma}, semi-classical quantum effects \cite{quantum}, dark matter \cite{dark matter} and the cosmological expansion \cite{cosmic-shadow-2}. Motivated by the accelerated expansion of the universe, we seek to investigate the effect of the cosmological constant on the shadow shape.

Dark energy, one of the most important problems in cosmology, is assumed to drive the current cosmic accelerated expansion. The cosmological constant, $ \Lambda $, is among the favored candidates responsible for this acceleration. The cosmological constant has imprints on various phenomena in structure formation and astrophysics. These effects can appear as the integrated Sachs-Wolfe effect in the Cosmic Microwave Background and large scale structure observables. Moreover, the cosmological constant can change the luminosity distance in the gravitational wave observations. Recently, the role of the cosmological constant in gravitational lensing has been the subject of focused studies \cite{Islam:1983rxp,Rindler:2007zz,Piattella:2015xga,Butcher:2016yrs,Faraoni:2016wae,Lebedev:2016kun,He:2017alg,Zhao:2015fya,Ishihara:2016vdc}. It seems there is no general agreement on the final results. The black hole shadow arises as a result of gravitational lensing in strong gravity regime. Thus, one can investigate the influence of the cosmological constant on the shadow of black holes. The first attempt to derive the shadow size as measured
by the comoving observers was performed by Perlick et.al \cite{cosmic-shadow-1} using the Kottler metric. Authors in \cite{cosmic-shadow-2}  presented an approximate method based on the angular size-redshift relation for the general case of any multicomponent universe. Although the cosmological constant correction on the shadow shape of black holes has been studied in \cite{Grenzebach:2014fha,Zakharov:2014pca,Nitta:2011in, Mann}, choosing the appropriate (cosmological) observers is another issue that one has to consider.

To study the effect of the cosmological constant in the lensing of a cosmological structure or shadow of a black hole, one has to model a cosmological structure like a black hole in the expanding background. There have been many attempts to model a cosmological black hole (CBH). The authors in \cite{cbh,man} have studied the effect of the expanding background. Since a cosmological structure like a galaxy, a cluster of galaxies or a supermassive black hole in the center of a quasar have dynamical evolution due to their dynamics or cosmological expansion, one has to use a dynamical cosmological model to study its lensing \cite{Mood:2013uba}. 
It's been shown \cite{man} that the background expansion for CBHs leads to the formation of voids. These voids are formed between the black hole and the expanding background and prevent the black hole's matter flux from increasing. After this phase of the black hole evolution, the black hole can be approximated as a point mass \cite{tohid}. As a result, the final stage of evolution can be approximated by a point mass located in a dark energy dominated background.

There is a well-known solution of the Einstein equation for a point mass black hole in the presence of the cosmological constant known as Schwarzschild-de Sitter/Kottler spacetime. This solution is written in static coordinates.
To find the geodesic observers we have to apply an appropriate coordinate transformation to present the  Schwarzschild-de Sitter/Kottler spacetime in the cosmological synchronous gauge \cite{Podolsky:1999ts,tohid}.

This paper is organized as follows.  In section II we introduce a point mass cosmological black hole. In section III,  we calculate the angular size of the shadow in the Schwarzschild-de Sitter spacetime as measured by a cosmological comoving observer and calculate the different limits.  The conclusion and discussion are given in Section IV.\\

%%%%%%%%%%%%%%%%%%%%%%%%%%%%%%%%%%%%%%%%%%%%
\section{Point mass black holes metrics}
%%%%%%%%%%%%%%%%%%%%%%%%%%%%%%%%%%%%%%%%%%%%

It is known that the Schwarzschild-de Sitter is a standard metric for the CBH, but Schwarzschild-de Sitter has been written in the static coordinates. To study the point mass metric in an expanding background, one needs to find the Schwarzschild-de Sitter metric in the cosmological coordinates.
Since the standard cosmological metrics are written in the synchronous coordinate, one has to transform the Schwarzschild-de Sitter to the synchronous coordinate \cite{Podolsky:1999ts,tohid}.
The  Schwarzschild-de Sitter metric in the static coordinates is given by 
\begin{equation}
\label{a1}
ds^{2} =  - \Phi   dt^{2}  + \Phi ^{-1}  dR^{2} + R^{2}\, d \Omega ^{2},
\end{equation}

where

\begin{equation}
\label{s-d}
\Phi =   1- \frac{\Lambda}{3} R^{2} - \frac{2M}{R}.
\end{equation} 
In the weak field limit, one can interpret $\frac{\Lambda}{3} R^{2}$ as the cosmological constant potential and compare it with the gravitational potential of the black hole, $M/R$ for the gravitational interactions. Using coordinate transformations  

\begin{equation}
\label{a2}
\begin{split}
&  d \tau = dt - \frac{\sqrt{1 - \Phi}}{\Phi}dR,
\\& dr = -dt +  \frac{dR}{\Phi \sqrt{1- \Phi}} ,
\end{split}
\end{equation}
metric~\eqref{a1}  will be written as

\begin{equation}
\label{comoving-coordinate }
ds^{2} = -d \tau ^{2} +\, \left( \frac{2M}{R} +  \frac{\Lambda}{3}R^{2}\right) dr^{2} + R^{2}d\Omega^{2}.
\end{equation}
We need to find $R$ as a function of $(r,\tau)$.
Using the coordinate transformation~\eqref{a2} yields
\begin{equation}
\label{t+r }
\tau + r = \int \frac{dR}{\sqrt{\frac{\Lambda}{3}R^{2} + \frac{2M}{R}}} =  \frac{2}{\sqrt{3 \Lambda}} \ln\,\left(  \frac{\Lambda R^{\frac{3}{2}} + \sqrt{6M \Lambda + \Lambda^{2}R^{3}}}{\kappa}\right).
\end{equation}
Solving Eq.~\eqref{t+r } for R, we get
\begin{equation}
\label{Req}
R=   \frac{{\rm e}^{ - \sqrt{\frac{\Lambda}{3} } (r + \tau) }\, \left( {\rm e}^{\sqrt{3 \Lambda} (r+ \tau ) } \kappa^2-6\Lambda M\right) ^{\frac{2}{3}}   }{(2 \kappa \Lambda)^{\frac{2}{3}}},
\end{equation}
where $\kappa$ is a constant. When $M=0$, one can define a new radial coordinate as
\begin{align}
\tilde{R}=\left( \frac{\kappa}{\Lambda}\right)^{\frac{2}{3}}{\rm e}^{\sqrt{\frac{\Lambda}{3}}r}. 
\end{align}
In this coordinate we obtain the de Sitter metric in the synchronous coordinates as
\begin{align}
ds^2=-d\tau^2+a(\tau )^2\left(d\tilde{R}^2+\tilde{R}^2\,d\Omega^2 \right), 
\end{align}
where $a(\tau)={\rm e}^{\sqrt{\frac{\Lambda}{3}}\tau}$.
 The four-velocity of the comoving observer, $dr=0$, who measures the proper time $\tau$ can be written using  transformation~\eqref{a2} in the static coordinates  as
\be
\label{U-def}
U^\mu=(\frac{1}{\Phi},\sqrt{1-\Phi},0,0),
\ee
Four-acceleration of this observer is zero. Hence, it is called the comoving cosmological observer. In contrast to the metric introduced in \cite{cosmic-shadow-1}, the metric~\eqref{comoving-coordinate } is the appropriate metric in which \emph {r=constant} world lines are attached to the geodesic (comoving) observers.
The position of each comoving observer $r=contant$ in static coordinates is given by Eq.~\eqref{Req}.
In the next section, we calculate the angular shadow size seen by the cosmological comoving observers. \\

%%%%%%%%%%%%%%%%%%%%%%%%%%%%%%%%%%%%%%%%%%%%
\section{Shadow for cosmological comoving observer}
%%%%%%%%%%%%%%%%%%%%%%%%%%%%%%%%%%%%%%%%%%%%

Before calculating the angular  size of the shadow for the cosmological observers, let us present the result for  the static observers from \cite{cosmic-shadow-1}. The shadow   is locally seen by a static observer at 
a spacetime point $(t_O,R_O,\theta_O = \pi /2, \phi _O =0)$  outside the black hole horizon and inside the cosmological horizon \footnote{For the central observer which is located in center of a spherical coordinate, the current value of cosmological constant with astrophysical black holes mass gives two horizons for the Schwarzschild-de Sitter metric. One horizon is known as the black hole horizon and the other is known as the cosmological horizon. Each observer in de Sitter spacetime has its own cosmological horizon. The cosmological observer (who see the black hole ray in the center) is located between two horizons of the central observer in the static coordinates. Note that, no light from a source outside the cosmological horizon can  get to the central observer which is located in center of the cosmological horizon.}. The Lagrangian  for geodesics in the equatorial plane gives 
\begin{equation}\label{eq:Lagr}
\mathcal{L} (x , \dot{x} ) \, = \, 
\dfrac{1}{2} \, \left(- \, \Phi(R)  \dot{t}{}^2
+ \dfrac{\dot{R}{}^2}{\Phi(R)} + R^2 \dot{\phi}{}^2 \right) 
\,.
\end{equation}
The Euler-Lagrange equations for $t$ and $\phi$ coordinates
give  two constants of motion,
\begin{equation}\label{eq:com}
E \, = \, \Phi(R) \, \dot{t} \, , \qquad
L \, = \, R^2 \,  \dot{\phi}
\end{equation}
where $E$ and $L$ represent the energy and angular momentum of the geodesic.
For null geodesics we have 
\begin{equation}\label{eq:L=0}
- \,  \Phi(R) \, \dot{t}{}^2 \, + \,
\dfrac{\dot{R}{}^2}{\Phi(R)} \, + \, R^2 \dot{\phi}{}^2
\, = \, 0 \, .
\end{equation}
One can solve $(\dot{R}/\dot{\phi})^2 = (dR/d \phi )^2$ by using Eq.~\eqref{eq:com}
to get the null geodesics orbits as
\begin{equation}\label{eq:ol}
\left( \dfrac{d R}{d \phi} \right) ^2 \, = \, R^4 \left(  
\dfrac{E^2 }{ L^2} + \dfrac{\Lambda}{3} \, - \, 
\dfrac{1}{R^2} \, + \,\dfrac{2M}{R^3} \right) \, .
\end{equation}
\\

If we apply the conditions $d R/d\phi =0$ and $d^2 R/d\phi ^2 =0$, we 
find that there is a circular null geodesic orbit at radius $R = 3M$ where the  constants of motion for this circular null geodesic satisfy
\begin{equation}\label{eq-shadow}
\dfrac{E^2}{L^2} \, = \, \dfrac{1}{27 \, M^2} \, - \, \dfrac{\Lambda}{3} \, . 
\end{equation}

\begin{figure}[htbp!]
    \centering \includegraphics[width=0.8 \columnwidth]{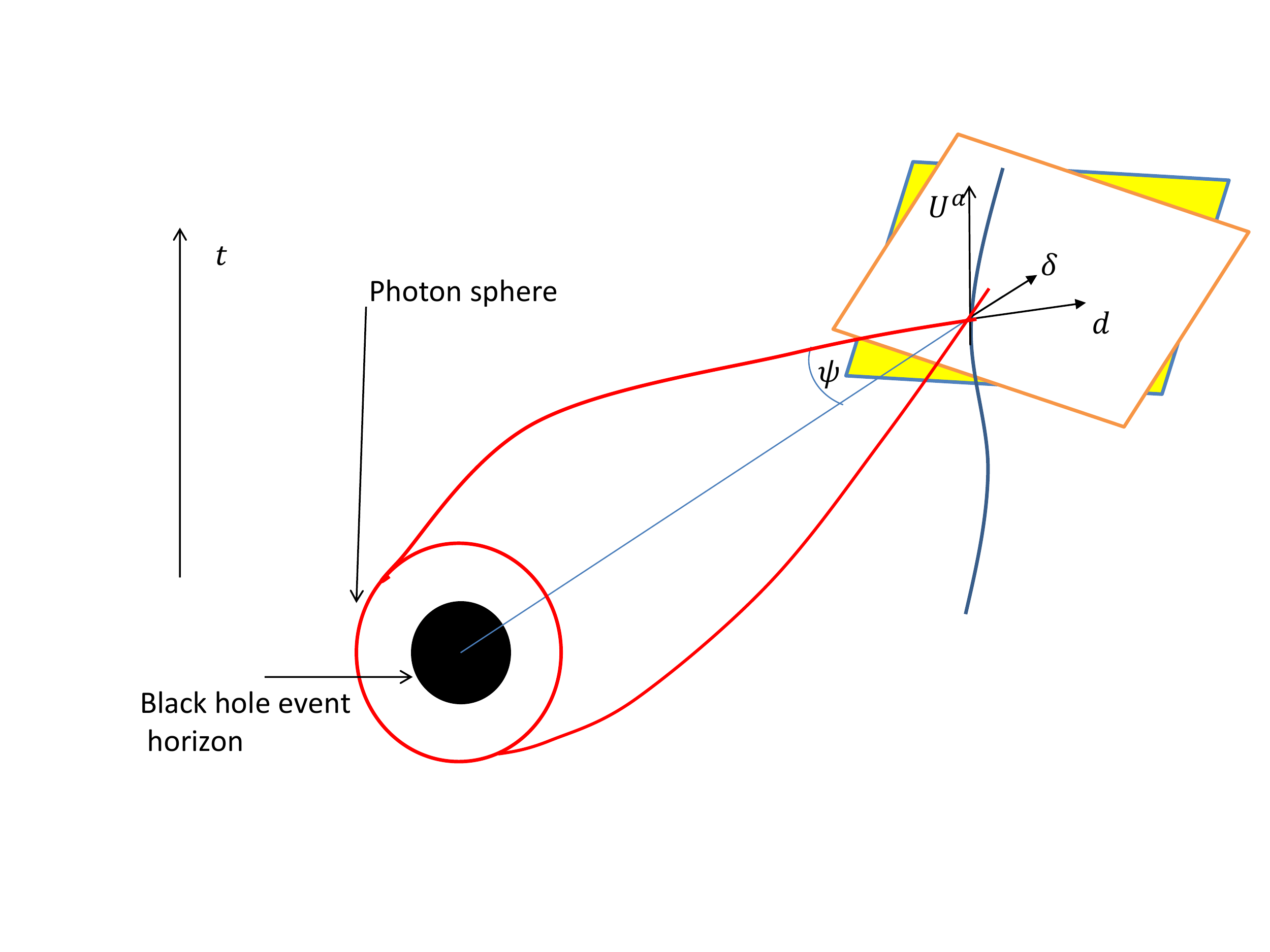}  \caption{Black hole shadow relative to the cosmological comoving observer. $\delta$ is the observer's position vector and $d$ represent the vector tangent to the photon path. $U^\alpha$ is the comoving observer velocity. The static observer measures $\psi$.}     \label{fig:1}    
\end{figure}

The boundary of the shadow is determined by the initial directions of last light rays that asymptotically spiral towards the outermost photon sphere (see Fig (\ref{fig:1}) ). Using Eq.~\eqref{eq-shadow}, we find that $\dfrac{d R}{d \phi}$  for this light ray is given by
\be
\label{shdow-light}
( \dfrac{d R}{d \phi}) ^2  = \, R^4 \left(  
\dfrac{1 }{27 M^2}  - \, 
\dfrac{1}{R^2} + \,\dfrac{2M}{R^3} \right),
\ee
As shown in Fig. \eqref{fig:1} the angular radius of the shadow is $ 2\psi  $.  Since the comoving observer measures the shadow angle in his rest frame which is orthogonal to his four-velocity vector $U^{\mu}$, we have to build the projection onto his rest frame as
\be
h_{\mu \nu}=g_{\mu \nu}+U_\mu U_\nu
\ee
We consider a null geodesic in the $\theta=\frac{\pi}{2}$ plan.
The observed shadow angle is given by the inner product of the observer position vector denoted by $\delta^\mu$ and the tangent to the photon trajectory denoted by $d$.
In these coordinates, $\delta^\mu=(0,R_o,0,0)$ and $d^\mu=k^\mu=\frac{L}{R^2} \left(\frac{ER^2}{L\Phi},A,0,1 \right) $ where $A=\frac{d R}{d \phi}$ on the null geodesics.
As Ishak and Rindler have shown \cite{Rindler:2007zz}, one can calculate the bending angle using the invariant formula for the cosine of the angles between two  directions $ d $ and $ \delta $. This gives
\be
\cos \psi=\dfrac{A}{\sqrt{A^2+\Phi R^2}},
\ee
where dot denotes the inner product. For the static observers of the Schwartzchild-de Sitter spacetime, one can show
\begin{align}
\sin^2 \psi=\frac{\left( 1-\frac{2 M}{R}-\frac{\Lambda}{3}R^2\right) }{ \frac{R^2}{27 \, M^2} \, - \, \frac{\Lambda}{3} R^2}
\end{align}
\cite{cosmic-shadow-1}. As stated, each observer measures the angles in his rest frame. To calculate the shadow angular size relative to the cosmological comoving observer,  we have to project these two vectors to his rest frame which is perpendicular to the $U^\mu$ as
\be
d_U^\mu=h^{\mu \nu} d_\nu ,~~\delta_U^\mu=h^{\mu \nu} \delta_\nu.
\ee
It's been shown that the projection effects are important in measuring cosmological observables \cite{cosmological observable}. The angle measured with respect to this observer  is given by
\begin{align}
\cos \psi_U=\dfrac{d_U.\delta_U}{|d_U||\delta_U|}=\frac{d^\mu \delta_\mu+\left( d.U\right) \left(\delta.U \right) }{ d.U \sqrt{\delta^\mu \delta_\mu+(\delta.U)^2}  }.
\end{align}
Expanding this equation  and using $g_{rr}=\frac{1}{\Phi}$ and $g_{\phi \phi}= R^2$ , we get
\begin{align}
&d^\mu\delta_{\mu}=\frac{LA}{R\Phi},\\
&d.U=-\frac{E}{\Phi}+\frac{LA\sqrt{1-\Phi}}{R^2\Phi},\\
&\delta.U=\frac{R\sqrt{1-\Phi}}{\Phi},
\end{align}
and
\ba
\cos \psi_U=\dfrac{\frac{A}{R\Phi}+\left( -\frac{E}{L\Phi}+\frac{A\sqrt{1-\Phi}}{R^2\Phi}\right) \left(\frac{R\sqrt{1-\Phi}}{\Phi} \right) }{\left(-\frac{E}{L\Phi}+\frac{A\sqrt{1-\Phi}}{R^2\Phi} \right)\sqrt{\frac{R^2}{\Phi}+\frac{R^2(1-\Phi)}{\Phi^2}} }
\label{shad-ang}        
\ea
where for  photons starting on the photon sphere $ \left(  \dfrac{d R}{d\phi}\right)   = - \, R^2 \sqrt{ 
    \frac{1 }{27 M^2}  - \, 
    \frac{1}{R^2} + \,\frac{2M}{R^3} } $. Note that the observed angle is calculated at the observer position, $R_O$. The Eq.~\eqref{shad-ang} relates the angle measured by the static observers to the comoving observers. Finally,
\begin{align}
\cos\psi_U=\dfrac{\Phi R\sqrt{A^2+R^2\Phi}\cos\psi+R\sqrt{1-\Phi}\left(-\frac{ER^2}{L}+A\sqrt{1-\Phi} \right) }{R\left(-\frac{ER^2}{L}+A\sqrt{1-\Phi} \right) }.
\end{align}

The black hole horizon and cosmological horizon in Schwarzschild-de Sitter spacetimes are located where $\Phi(R)=0$. The observer at the black hole horizon, where we have $\Phi=0$, sees $\psi_U=\pi$.
 For an observer on the black hole photon sphere $R= 3 M $, we have $\psi=\pi/2$. As a result we find that
\begin{align}
\cos\psi_U=\sqrt{\frac{2}{3}+3 M^2\Lambda}. 
\end{align}
If the observer is located on the cosmological horizon, we have $\Phi=0$. 

Using the equation \eqref{eq-shadow} and \eqref{shdow-light} one can get
\be
A^2=\frac{R^4 E^2}{L^2}-R^2 \Phi.
\ee
On the horizon, we have $A=\pm \frac{R^2 E}{L}$.
The positive sign is not acceptable because using the current value of cosmological constant where $\lambda M^2 <1$, the value, $\cos\psi_U$, in \eqref{shad-ang} get $|\cos\psi_U|>1$. Hence,  $A= - \frac{R^2 E}{L}$. 
 
 In this case, we find that $\psi_U=0$.
Our result differs from the result in \cite{cosmic-shadow-1} because they used a different observer.

\begin{figure}[htbp!]
   \centering    \includegraphics[width=0.7 \columnwidth]{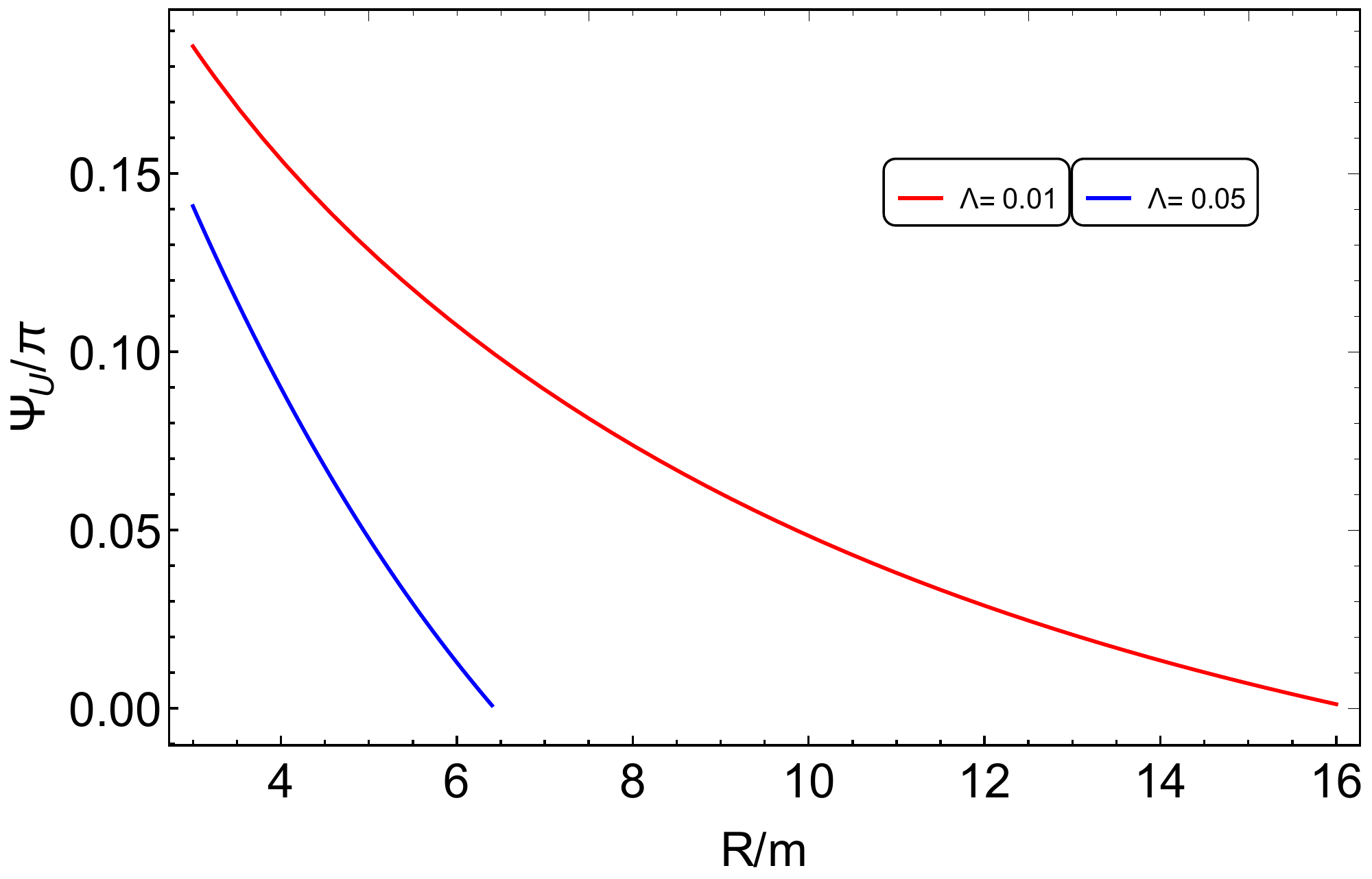}
    \caption{The observed shadow angle ($\psi_U$) for the comoving observer as a function of distance. The blue curve is for $\Lambda=0.05$ and the red curve for $\Lambda=0.01$. The shadow angle decreases faster for the larger value of the cosmological constant. On the horizon, we have $\psi_U=0$. We have set $M=1$. }
    \label{fig2}
\end{figure}

To illustrate the cosmological constant effect on the shadow angle we have plotted the shadow angle as a function of distance for two different values of the cosmological constant in Fig.~\ref{fig2}. We have used $M=1$ in the calculations. One can see that the bigger value for the cosmological constant decreases the shadow angle size. This result agrees with the result in \cite{cosmic-shadow-2}. It is because the cosmological constant increases the angular diameter distance.\\

In order to estimate the cosmological constant effect on the shadow angle, let us calculate the shadow angle for the black hole in M87. If we evaluate the Eq. \eqref{shad-ang}  for the black hole M87 with mass, $6.5   \times    10^9 M_{\odot}$, and distance, $R=16.4 Mpc$, we get
\be
\cos\psi_U \sim\, \left( 1+\mathcal{O}(10^{-7})\right)   \cos \psi.
\ee
Given that the current resolution for present-day VLBI technology allows only to resolve angles of a few dozen microarcseconds, we don't need to consider the projection effects to interpret the shadow observations. This is true as long as our distance to the black hole is less than $R_O ~\Lambda^2 \ll 10^{-2}$, which gives a small value of the cosmological constant potential. \\

There are various black hole candidates observed so far. We know their estimated mass and distance. We calculate their shadow angular size in microarcseconds and also estimate the effect of cosmological constant for various candidates. In table \ref{1} we present a catalog of the supermassive black holes with the known mass and distance. We give their shadow angular size with and without the cosmological constant. We sort the data in terms of the shadow size fractional difference, $\frac{\delta \psi}{\psi}=\frac{\psi_{\Lambda}-\psi_{\Lambda=0}}{\psi_{\Lambda}}$.  The cosmological constant effect is considerable on large cosmological distances. The effect could be $30\;precent$ for far away sources. Improving the angular resolution in Event horizon telescope for distance black holes, can be used to measure the cosmological constant value (Hubble constant).\\

\begin{table}[b]
    \caption{ supermassive black holes with known mass and distance. In this table, we compare the effect of the cosmological constant on their shadow size in microarcseconds. The fractional difference could be of order $30~ precent$ for the most distance supermassive black holes. }
    \label{1}
    \begin{tabular}{l|l|l|l|l|l}
        Name                                                                              & Mass ($M_{\odot}=1$) & Distance         & Shadow  size without $\Lambda$  & Shadow  size with $\Lambda$  & $\frac{\delta \psi}{\psi}$ \\ \hline  
        ULAS J1120+0641                                                                   & 2   $\times $   10\textasciicircum{}9             & 4.0 Gpc          & 0.0484855                                          & 0.32288                                         & 0.33                              \\
        ULAS J1342+0928                                                                   & 8    $\times $  10\textasciicircum{}8              & 4.00 Gpc         & 0.193942                                           & 0.129155                                        & 0.33                              \\
        Q0906+6930                                                                        & 2   $\times  $ 10\textasciicircum{}9              & 3.77 Gpc         & 0.0514435                                          & 0.0365801                                       & 0.28                              \\
        S5 0014+81                                                                        & 4  $\times $ 10\textasciicircum{}10             & 3.7 Gpc          & 1.04833                                            & 0.758733                                        & 0.27                              \\
        APM 08279+5255                                                                    & 2.3  $\times $ 10\textasciicircum{}10           & 3.69 Gpc         & 0.6195                                             & 0.4591                                          & 0.25                              \\
        TON 618                                                                           & 6.6  $\times $ 10\textasciicircum{}10           & 3.18 Gpc         & 2.0126                                             & 1.62044                                         & 0.19                              \\
        Phoenix   Cluster BH & 2  $\times $ 10\textasciicircum{}10             & 1.74 Gpc         & 1.11461                                            & 1.05429                                         & 0.05                              \\
        OJ 287 primary                                                                    & 1.8   $\times $  10\textasciicircum{}10           & 1.073 Gpc      & 1.63129                                            & 1.59848                                         & 0.02                              \\
        Abell 1201 BCG                                                                    & 1.3    $\times $  10\textasciicircum{}10     & 838.5 Mpc        & 1.50432                                            & 1.48584                                         & 0.012                             \\
        H1821+643                                                                         & 3  $\times $ 10\textasciicircum{}10             & 1.0 Gpc          & 2.90913                                            & 2.85809                                         & 0.01                              \\
        3C 273                                                                            & 8.86   $\times $ 10\textasciicircum{}8    & 749 Mpc          & 0.113931                                           & 0.112814                                        & 0.0098                            \\
        Hercules A (3C 348)                                                               & 4  $\times $ 10\textasciicircum{}9              & 643.9 Mpc        & 0.603241                                           & 0.598887                                        & 0.0072                            \\
        IC 1101                                                                           & 4  $\times $ 10\textasciicircum{}10        & 320.4   Mpc & 24.2427                                            & 24.1995                                         & 0.0017                            \\
        Cygnus A                                                                          & 1  $\times $ 10\textasciicircum{}9              & 232 Mpc          & 0.417978                                           & 0.417587                                        & 0.00093                           \\
        Holmberg 15A                                                                      & 4.0   $\times $ 10\textasciicircum{}10     & 216 Mpc          & 17.9576                                            & 17.943                                          & 0.00081                           \\
        RX J124236.9-111935                                                               & 1  $\times $ 10\textasciicircum{}8              & 200 Mpc          & 0.0484855                                          & 0.0484517                                       & 0.00069                           \\
        NGC 6166                                                                          & 3  $\times $ 10\textasciicircum{}10             & 142 Mpc          & 20.4868                                            & 20.4796                                         & 0.00035                           \\
        NGC 6166                                                                          & 1  $\times $ 10\textasciicircum{}9              & 142 Mpc          & 0.682894                                           & 0.682655                                        & 0.00035                           \\
        Markarian 501                                                                     & 9  $\times $ 10\textasciicircum{}9              & 140 Mpc          & 6.23385                                            & 6.23172                                         & 0.00034                           \\
        NGC 3842                                                                          & 9.7  $\times $ 10\textasciicircum{}9            & 99.6 Mpc         & 9.44396                                            & 9.44233                                         & 0.00016                           \\
        NGC 4889                                                                          & 2.1   $\times $ 10\textasciicircum{}10     & 94.43   Mpc & 21.6637                                            & 21.6604                                         & 0.00015                           \\
        NGC 1270                                                                          & 1.2  $\times $ 10\textasciicircum{}10           & 78 Mpc           & 14.9186                                            & 14.917                                          & 0.0001                            \\
        NGC 1271                                                                          & 3.0  $\times $ 10\textasciicircum{}9            & 76.3 Mpc         & 3.8278                                             & 3.82742                                         & 0.0001                            \\
        NGC 5548                                                                          & 6.71  $\times $ 10\textasciicircum{}7    & 75.01 Mpc        & 0.086627                                           & 0.086618                                        & 0.000097                          \\
        NGC 1277                                                                          & 1.2  $\times $ 10\textasciicircum{}9            & 73 Mpc           & 1.59404                                            & 1.5939                                          & 0.000092                          \\
        NGC 1275                                                                          & 3.4  $\times $ 10\textasciicircum{}8            & 68.2 Mpc         & 0.484855                                           & 0.484816                                        & 0.00008                           \\
        NGC 1281                                                                          & 1  $\times $ 10\textasciicircum{}10             & 60   Mpc       & 16.1618                                            & 16.1608                                         & 0.000062                          \\
        NGC 1600                                                                          & 1.7   $\times $ 10\textasciicircum{}10    & 45.6 Mpc         & 36.1515                                            & 36.1501                                         & 0.00003                           \\
        NGC 3783                                                                          & 2.98   $\times $ 10\textasciicircum{}7    & 41.60 Mpc        & 0.0709544                                          & 0.0709523                                       & 0.000029                          \\
        NGC 4261                                                                          & 4  $\times $ 10\textasciicircum{}8              & 29.4   Mpc   & 1.31933                                            & 1.31931                                         & 0.000015                          \\
        NGC 3227                                                                          & 4.22   $\times $ 10\textasciicircum{}7    & 24  Mpc       & 0.150844                                           & 0.150842                                        & 0.000012                          \\
        NGC 1399                                                                          & 5  $\times $ 10\textasciicircum{}8              & 20.23 Mpc        & 2.42427                                            & 2.42426                                         & 0.0000069                         \\
        NGC 7469                                                                          & 12.2   $\times $ 10\textasciicircum{}6     & 20.1 Mpc         & 0.0581826                                          & 0.0581822                                       & 0.0000069                         \\
        Messier 58                                                                        & 7  $\times $ 10\textasciicircum{}7              & 19.1 Mpc         & 0.357261                                           & 0.357259                                        & 0.0000062                         \\
        NGC 4151 primary                                                                  & 4  $\times $ 10\textasciicircum{}7              & 19 Mpc           & 0.204149                                           & 0.204148                                        & 0.0000062                         \\
        Messier 85                                                                        & 1  $\times $ 10\textasciicircum{}8              & 18.5   Mpc   & 0.524167                                           & 0.524164                                        & 0.0000059                         \\
        Messier 60                                                                        & 4.5   $\times $ 10\textasciicircum{}9      & 17.38 Mpc        & 25.1076                                            & 25.2074                                         & 0.000005                          \\
        Messier 49                                                                        & 5.6  $\times $ 10\textasciicircum{}8            & 17.14   Mpc & 3.19434                                            & 3.19432                                         & 0.000005                          \\
        Messier 84                                                                        & 1.5  $\times $ 10\textasciicircum{}9            & 16.83 Mpc        & 8.64269                                            & 8.64265                                         & 0.0000049                         \\
        M60-UCD1                                                                          & 2  $\times $ 10\textasciicircum{}7              & 16.5 Mpc         & 0.117541                                           & 0.11754                                         & 0.0000047                         \\
        Messier 61                                                                        & 5  $\times $ 10\textasciicircum{}6              & 16.10   Mpc & 0.303034                                           & 0.303033                                        & 0.0000044                         \\
        Messier 108                                                                       & 2.4  $\times $ 10\textasciicircum{}7            & 14.01 Mpc        & 0.166236                                           & 0.166235                                        & 0.0000034                         \\
        Messier 59                                                                        & 2.7  $\times $ 10\textasciicircum{}8            & 15.35 Mpc        & 1.74548                                            & 1.74547                                         & 0.0000032                         \\
        Messier 105                                                                       & 2  $\times $ 10\textasciicircum{}8              & 11.22 Mpc        & 1.76311                                            & 1.7631                                          & 0.0000021                         \\
        NGC 3115                                                                          & 2  $\times $ 10\textasciicircum{}9              & 9.7   Mpc    & 19.994                                             & 19.994                                          & 0.0000016                         \\
        Sombrero Galaxy                                                                   & 1  $\times $ 10\textasciicircum{}9              & 9.55   Mpc  & 10.154                                             & 10.154                                          & 0.0000015                         \\
        Messier 96                                                                        & 48000000                             & 9.6   Mpc    & 0.489959                                           & 0.489958                                        & 0.0000015                         \\
        Messier 87                                                                        & 6.77   $\times $ 10\textasciicircum{}9          & 16.4   Mpc   & 40.0301                                            & 40.0299                                         & 0.000001                          \\
        Centaurus A                                                                       & 5.5  $\times $ 10\textasciicircum{}7            & 3–5 Mpc          & 1.33335                                            & 1.33335                                         & 0.00000027                        \\
        Messier 82 (Cigar Galaxy)                                                         & 3  $\times $ 10\textasciicircum{}7              & 3.5–3.8 Mpc      & 0.83118                                            & 0.83118                                         & 0.00000021                        \\
        Andromeda Galaxy                                                                  & 2.3  $\times $ 10\textasciicircum{}8            & 778   kpc     & 28.6675                                            & 28.6675                                         & 0.00000001                        \\
        Messier 32                                                                        & 5  $\times $ 10\textasciicircum{}6              & 763   kpc     & 0.69365                                            & 0.69265                                         & 8.5E-09                           \\
        Sagittarius A*                                                                    & 4.3  $\times $ 10\textasciicircum{}6            & 7860  pc      & 53.4584                                            & 53.4584                                         & 1E-12                            
    \end{tabular}
\end{table}

%%%%%%%%%%%%%%%%%%%%%%%%%%%%%%%%%%%%%%%%%%%%%%%%%%%%%%%%%%
\section{Conclusion and discussion}
%%%%%%%%%%%%%%%%%%%%%%%%%%%%%%%%%%%%%%%%%%%%%%%%%%%%%%%%%%
The existence of the cosmological constant changes the angular size of the black hole shadow for the comoving cosmological observers. 
Our work provides a possible tool to calculate the effect of the cosmological constant on the shadow size. Projecting the shadow angle to this observer rest frame, we calculated the shadow size observed by the comoving cosmological observers.
In accordance with this article analysis, comoving observers see the shadow angular size $\psi_U=\arccos\left(\sqrt{\frac{2}{3}+3 M^2\Lambda} \right) $  on the black hole photon sphere at $R=3 M$. As the observer moves away from the black hole horizon towards the cosmological horizon, the shadow angle decreases to $\psi_U=0$ on the cosmological horizon. Our result differs from the result in the reference \cite{cosmic-shadow-1} because we used the synchronous form of the metric \eqref{comoving-coordinate } in which $r=constant$ world lines are geodesics.
 
Using the mass and distance value for the supermassive  black hole M87, one gets $\cos\psi_U \sim \left( 1+\mathcal{O}(10^{-7})\right)  \cos \psi$. The current resolution for the present-day VLBI technology allows only to resolve the angles of a few dozen microarcseconds. As a result,  we don't need to consider the projection effects to interpret the shadow observations. This is true as long as our distance to the black hole is less than $R_O ~\Lambda^2 \ll 10^{-2}$ which gives a small value of the cosmological constant potential. In the last part, we gave a catalog of the supermassive black holes with the known mass and distance. We compared their shadow angular size with and without the cosmological constant. The effect of the cosmological constant is observable on the shadow size on large distances. We find the effect of cosmological constant could be $3\;precent$ for distant known sources like the Phoenix Cluster supermassive black hole

%%%%%%%%%%%%%%%%%%%%%%%%%%%%%%%%%%%%%%%%%%%%%%%%%%%%%%%%

\end{document}